# Students' Perceptions of the Effectiveness of Discussion Boards:

What can we get from our students for a freebie point?

Abdel-Hameed A. Badawy
Electrical and Computer Engineering Department,
University of Maryland,
College Park, MD, USA

*Abstract*— **In this paper, we investigate how the students think of their experience in a junior (300 level) computer science course that uses blackboard as the underlying course management system. Blackboard's discussion boards are heavily used for programming project support and to foster cooperation among students to answer their questions/concerns. A survey is conducted through blackboard as a voluntary quiz and the student who participated were given a participation point for their effort. The results and the participation were very interesting. We obtained statistics from the answers to the questions. The students also have given us feedback in the form of comments to all questions except for two only. The students have shown understanding, maturity and willingness to participate in pedagogy-enhancing endeavors with the premise that it might help their education and others' education as well.**

*Keywords- collaborative learning; cooperative learning; peer learning; teaching evaluation; pedagogy; discussion boards; Blackboard; pedagogy; survey; student feedback.[1]*

## I. INTRODUCTION

The students in both computer science and computer engineering have to take a C language [1] programming course that is worth four credits, which involves a lot of programming. This course requires a lot of effort from the teaching staff to help the students with their debugging issues and to clarify ambiguous, unclear or even mistakes in the requirements of the assigned programming projects. This c programming course has a password-protected website where lecture slides, project descriptions and other related documents can be retrieved. One of the courses that follow the C programming course is a regular three credits course that teaches computer organization, which also has a significant programming component in it as well. In some cases, projects can be as large as several hundreds of lines of code. Course management software is used in the management of the follow-up course. The course management software adopted by the University of Maryland at College Park (UMD) is called Enterprise Learning Management System (ELMS) [3] and is powered by Blackboard© version 8.0 at the time of conducting this study. Blackboard© version 9.1 was released recently [2]. We conducted a simple survey study to assess the perception of the students of the follow-up course to the use of the discussion boards of ELMS.

---

[1] A shorter version of this paper appeared in [4, 17].

## II. RELATED WORK

The volume of work, in the last few years in education and the theory of teaching and learning, is beyond any single person ability to follow. Many conferences and journals are concerned with Teaching and Learning. It is clear to us that we cannot be very thorough in our coverage of all related work to this work; nevertheless, we will try here to touch on some of the seminal works in the areas related to our work.

In terms of pedagogy, cooperative learning and collaborative learning are the two most closely related pedagogies to our theme in this paper. A great chapter that introduces cooperative, collaborative and peer learning appears in Wilbert J. McKeachie's "Teaching Tips" book [11]. Collaborative learning and cooperative learning are sometimes used synonymously even though in the literature they are different. Cooper discussed the differences and similarities between collaborative and cooperative learning in [18]. Gokhale examined collaborative learning techniques in a study and has shown that collaboration among students enhances learning and increases critical thinking [9]. Felder discusses effective techniques and methodologies for using collaborative learning in teaching [20].

Discussion boards are a commonly available feature in many online course management systems such as Blackboard. Jeong, of Florida State University, has published extensively on topics related to discussion boards in general [12, 13, 14]. He investigated how can online discussion boards engage all students and promote interaction among them [12]. He examined facilitating online discussions effectively [14]. He also designed computer-based tools and methods to analyze discussion boards and give instructors methods to assess, grade and evaluate discussion boards. Northover investigated whether or not online discussion boards are a friend or a foe. He also suggested best practices to develop effective situations that can be easily delivered and assessed [15]. Dringus and Ellis used data mining techniques as an assessment strategy for evaluating discussion forums [16].

A large volume of work has been conducted to investigate the use and effectiveness of technology in the classroom and how do they influence student learning and the ways the professors are teaching. The investigated techniques range from presentation software such as Microsoft's PowerPoint®, course management software, course websites, online lecture notes, and discussion boards to personal electronic devices or what is coined as mobile learning or handheld learning [7, 8, 10, 11, 12, 15, 17, 19, 21, 22].

TABLE I. DISTRIBUTION OF ACCESSES TO EACH DISCUSSION FORUM.

| Forum | Accesses | % Total | Students | Staff | Ratio |
|---|---|---|---|---|---|
| Lab 1 | 4154 | 30% | 2756 | 1398 | 66% |
| Lab 2 | 3332 | 24% | 3241 | 91 | 97% |
| Lab 3 | 4557 | 33% | 3985 | 572 | 87% |
| Lab 4 | 876 | 6% | 876 | 0 | 100% |
| Technical | 833 | 7% | 700 | 133 | 84% |
| Total | 13752 | 100% | 11558 | 2194 | 84% |

TABLE II. MESSAGES DISTRIBUTION IN EACH DISCUSSION FORUM.

| Forum | Messages | % of Total | Students | Staff | Ratio |
|---|---|---|---|---|---|
| Lab 1 | 69 | 25% | 44 | 25 | 64% |
| Lab 2 | 73 | 27% | 69 | 4 | 95% |
| Lab 3 | 69 | 25% | 61 | 8 | 88% |
| Lab 4 | 32 | 12% | 32 | 0 | 100% |
| Technical | 31 | 11% | 24 | 7 | 77% |
| Total | 274 | 100% | 230 | 44 | 84% |

There exists a discussion board that students can use for instructor created topics. In case of the course under consideration in this work, each programming project had its own discussion forum. There were four projects (Labs 1, 2, 3 and 4). In addition, there was a forum for technical questions (called Technical) related to using the computer resources for the course. The traffic on the discussion boards for the first project was overwhelming. Therefore, we decided to design a survey to be able to get the feedback of the students about their experience with the discussion boards, its effectiveness and its contribution to their learning.

The rest of the paper is organized as follows: Section III introduces motivational statistics about number of messages accessed or generated. Section IV addresses the survey and its questions, the participation statistics and the logistics of conducting it and finally the results. Section V (Appendix) contains all the short answers the student gave for the open-ended question.

III. MOTIVATIONAL STATISTICS

ELMS (Blackboard) has built-in usage statistics collection tools and utilities that were very useful in our study. We have used these existing tools in ELMS for all parts of the courseware to see statistics like how many posts happened during the semester and how many times the posts were accessed, read or replied to.

Tables I and II respectively show the usage statistics of the forums in terms of total number of accesses to each forum and the number of unique posts and responses in each forum. In addition, we have broken up each of these according to the total number of events belonging to the students as opposed to the total number of forum posts belonging to the instructional staff.

Each of the tables shows six columns. The headings of the columns are exactly the same for both tables. The first column shows the name of the different forums. There were five forums in this course. There exists one forum per programming laboratory (totaling four programming labs) and a technical questions forum where the students would inquire about any connectivity or accessibility related questions to each other or to the instructional staff. The second column shows the number of accesses and messages per forum. Column three shows the percentage of the accesses and messages per forum with respect to the total number of accesses and messages to all the forums. Columns four and five show the distribution of each forum accesses and messages with respect to who viewed or wrote them whether it is the students or the instructional staff. Column six shows the percentage of the students' accesses and messages to the overall number of accesses and messages per forum.

Examining the reported numbers in tables I and II, one cannot ignore the very large number of accesses to the forums, which is almost 14K accesses relative to the relatively small number of messages exchanged on the forum of less than 300. Using some simple math, the ratio of the number of accesses per posted message is 50 to 1 *i.e.* the average per message views are 50. We need to keep in mind that the total number of students enrolled in this class was 65 students and there were four instructional staff for this course.

We get the following statistics:

1) On average, the number of messages posted per student is about three.
2) On average, the number of messages read per student is 178.
3) On average, the number of messages posted per staff member is 11.
4) On average, the number of messages read per staff member is 549.
5) The ratio of the number of messages posted by all the students to the messages posted by all the staff members is about five.
6) The ratio of the number of messages read by students to the messages read by all the staff members is about five.

We can conclude several conclusions from the above averages and ratios. First, on average, a student read far more messages than what he or she writes or posts.

TABLE III. SURVEY QUESTION (1).

| Do you think that ELMS is a helpful learning tool in this course? | % |
|---|---|
| Yes, it is a helpful learning tool. I love it. | 64% |
| No, it is not a useful learning tool. I hate it. | 2% |
| I neither love it nor hate. I am neutral. | 34% |

TABLE IV. SURVEY QUESTION (2).

| Do you prefer courses that use ELMS over other courses with a regular website? | % |
|---|---|
| Yes, courses are better with ELMS. | 63% |
| No, I like non-ELMS courses better. | 6% |
| It does not really matter. I do not care | 31% |

TABLE V. SURVEY QUESTION (3).

| How often do you post on the discussion boards? | % |
|---|---|
| Once Daily. | 2% |
| Once a week. | 19% |
| Once a month. | 21% |
| Once a semester. | 26% |
| Never posted. | 13% |
| I only read but I do not post. | 45% |

TABLE VI. SURVEY QUESTION (4).

| If you have questions, do you prefer to go to office hours or you try the boards first? | % |
|---|---|
| I prefer ELMS board posts. | 36% |
| I prefer to talk to someone face to face. | 43% |
| Depends on the time I have to figure out the answer. | 36% |
| I just ask a classmate. | 21% |

This means that with careful monitoring for the on-going flow of messages and posts on the forums we can reach the students with crucial clarifications and answers to hairy questions that we know that many of the students will read.

TABLE VII. SURVEY QUESTION (5).

| If you posted to the board and a fellow student answered your question, do you trust his answer? | % |
|---|---|
| Yes, sure I trust my classmates. | 53% |
| No, they might be wrong. | 9% |
| Only if someone from the instructional staff says it is a fine. | 26% |
| Not on all issues I trust my classmates' answers. | 21% |

TABLE VIII. SURVEY QUESTION (6).

| How many posts do you read? | % |
|---|---|
| I just read everything. | 40% |
| It is a waste of time. I read nothing. | 2% |
| I read posts depending on the title of the post. | 51% |
| I read posts related to my questions only. | 17% |

TABLE IX. SURVEY QUESTION (7).

| If there is a course with two sections one without an ELMS website and another with an ELMS website, which section would you enroll in? | % |
|---|---|
| The ELMS-based section. | 43% |
| The non-ELMS section. | 6% |
| It is not a factor at all. | 51% |

Another very important issue to notice here is that the staff members read three times more messages compared to the students, which makes perfect sense. The instructional staff is working hard to follow the discussions to make sure that the correct information is disseminated among the students and nobody is making some wrong, confusing, or misleading replies.

The statistics above suggest that the forums on ELMS are really a collaborative learning tool for the students. The students were the origin of most of the traffic on the forums. The students asked and replied to their own questions except in the rare cases where one of the instructional staff had to step in and correct or rectify a problematic issue. Clearly, the forums are a running archive for the students saving what the questions that were asked previously are and they were able to ask further questions. We can conclude here that the forums are a form of "student-directed online office hours" that are run by the students and monitored by the instructional staff.

TABLE X. SURVEY QUESTION (8).

| Currently you cannot submit anonymous questions and/or answers to the boards. If there was that option, would you participate more by reading and/or writing? | % |
|---|---|
| I would have participated more. | 30% |
| It would not matter to me. | 55% |
| I would not trust the posts if they were anonymous. | 15% |

TABLE XI. SURVEY QUESTION (9).

| Did you participate in this survey because of the freebie point? | % |
|---|---|
| Yes | 89% |
| No | 11% |

TABLE XII. NUMBER OF COMMENTS PER SURVEY QUESTION.

| Question | # of comments |
|---|---|
| Question 1 | 40 |
| Question 2 | 35 |
| Question 3 | 28 |
| Question 4 | 25 |
| Question 5 | 28 |
| Question 6 | 25 |
| Question 7 | N/A |
| Question 8 | N/A |
| Question 9 | 31 |

## IV. THE SURVEY, PARTICIPATION, LOGISTICS AND RESULTS

### A. Survey Participation and Logistics

The statistics we have shown so far show how much the forums helped the students. The total number of forum accesses and the other collected statistics are measures of the utility of the forum to the students. In order to increase the participation and reward the students who will participate in the survey we conducted, we gave each participating student a freebie point to be added to the total points a student scores in the course. Effectively, this point is 1% of the course grade.

Forty seven students completed the survey out of the 65 registered students for the class and 53 students attempted it, thus six students did not complete the survey they started. We consider this a great response from the students since it was a 72% completion rate and an 82% attempting rate. The goal at UMD for the end of semester course evaluations is 70% participation. The survey was open for participation for twenty-four hours only. The students of the class have scored a large participation turnout of over 90% in the campus-wide end of semester course evaluations, which clearly shows that we could have gotten better participation rate if the students were given more time to turn in the survey.

### B. Survey Results

Tables (III through XI) summarize the results that we have obtained from our survey. The first cell on the top left columns of each table is the question the table results are addressing. Table (XII) shows that we have asked the students to give us their own comments after every question except for questions seven and eight. The number of comments that we got were in many cases was very large. In many instances, we got very thoughtful statements. We will discuss some of the interesting comments. We are including all of the comments in the Appendix in section V.

### C. Discussion of the Results

In this subsection, we will go over some of the conclusions from the results reported in the tables. Table IV (Question 2) suggests that more than 60% of the students prefer ELMS over non-ELMS courses. Table V (Question 3) suggests that most of the students read the posts but do not write as much. Table VI (Question 4) suggest that ELMS is seen by students sometimes as a replacement of office hours. Table VII (Question 5) suggests that most of the students trust their fellow students for answering their questions. Table VIII (Question 6) suggests that about half the students read all posts and the other half reads posts depending on their titles. Table IX (Question 7) suggests that a very small percentage of the students prefer the non-ELMS sections, whereas close to half of the students prefer ELMS and a little above the half of the students do not bother whether the course is ELMS or not. Table X (Question 8) suggests that anonymity of the posts is not a factor for the students. Table XI (Question 9) suggest that most of the students found that the freebie point was a good push for them to take the survey.

### D. Limitations of the Study

This is study is speculative at many of its conclusions, yet there is plenty to draw from it. The sample size of the survey is small (around sixty students). We should have worded more tightly to get precise answers. For example, we should have asked how ELMS is useful to the students' learning. In some question, we allowed the students to check all that applied and thus in some questions the total of the percentage statistics is above 100%. This study should have happened along several years to remove any superfluous data or any jitter in the results. Two surveys should have been done in both courses of the sequence courses both of them not the junior level course only. Given all these limitations, we still think that the conclusion of the paper are relevant and useful for any computer science instructor who is teaching programming courses or courses with a heavy project and lab components.

*E. Contributions*

This paper aimed to articulate for faculty the importance of communication with the students. The students are willing to sacrifice their time to give us feedback. As faculty, our teaching style and pedagogy should be fully shaped and directed by the students. A continuous feedback loop should exist between student perceptions of different techniques and the adaptation of the technique(s). As faculty, we should disseminate our findings so that our fellow instructors can benefit from our experiences and experiments in the field of teaching and learning as well as in any other scientific discourse.

*F. Selected Student Comments*

Table XII summarizes the number of open-ended answers for the nine questions. The number of open-ended question range from 25 to 40 answers. We share in this section some of the student comments that are very thoughtful and interesting. Here are a selected set of the students' comments. Appendix I at the end of the paper contains a full listing of the students' comments organized per each survey question.

- "I do like giving feedback on pedagogy. Student feedback is hard to get/give in a big lecture hall. It benefits and improves the quality of the educational environment".
- "ELMS is alright, but if all the features were used, it could be (even) better".
- "I would have taken the survey without the point".
- "I also do feel that this survey is important".
- "Since there is no anonymity (in the boards), I am fairly certain no one would knowingly pass on false information (since their name associated with the post)".
- "If a teacher puts lecture notes and other materials on ELMS is helpful. Other teachers of mine has said they would use ELMS and then after 2 weeks they just gave up and just used emails".
- "Discussion boards are the most useful tool".
- "In the boards, I ask general questions to draw from the knowledge of the entire class rather than focusing all questions to TA's and the professor".
- "I think it would be helpful if all courses use ELMS".
- "(ELMS is) also nice for viewing grades and the syllabus and assignments".
- "Someone can post answers that are not perfect but still he/she knows better than I do".
- "If I can see the reasoning (in a classmate's answer of a question) and it makes sense, I can trust their answer well enough".
- "It (ELMS and its associated Discussion boards) makes classes without discussion sections easier to follow".
- "I do wish they would upgrade (the ELMS) GUI, it feels like a 90's web app".
- "Some classes do not use ELMS very well and others are very organized and utilize it".
- "(My issues with ELMS): you can't save your password to log into ELMS through google chrome which is painful".
- "(ELMS is) a great idea, but the interface is terrible".
- "When I have a question that is not answered by the discussion boards, I assume it is a question only I have and I will then go to a teacher to ask it".
- "Almost always, the face to face benefits are better than ELMS".
- "It is usually easier and more convient to post a message to the discussion board".

ACKNOWLEDGMENT

The author would like to acknowledge Dr. Michelle Hugue of the Computer Science Department, University of Maryland, College Park. The author also wants to acknowledge all the staff at the UMD's Center for Teaching Excellence (CTE) especially Henrike Lehnguth, Dr. Dave Eubanks and Prof. Spencer Benson. Without their feedback and support this study would not have materialized. Ms. Mary Maxson of the R. H. Smith School of Business' Information Technology for helpful feedback and suggesting to check a technology survey done at the School of Business in UMD. The author would also like to thank the anonymous reviewers for their feedback and insightful comments.

## V. APPENDIX I

We archive here the reflections of the students to each of the survey questions included in the survey. Some little editing occurred to maintain privacy and to hide names in some cases.

*1) Student Comments for Question 1*

The question was: Do you think that the ELMS website is a helpful learning tool in this course?

- Good for teachers who do not have their own web pages.
- A place for discussions and lecture notes is good.
- I think ELMS is very helpful for courses for the freshman or sophomore. But for a small class, I never use ELMS.
- I have found the discussion board to be the most useful tool on blackboard. It helps to be able to ask general questions to draw from the knowledge of the entire class rather than focusing all questions to TAs and the professor.
- I would say that I am neutral but I think that the discussion board is a very useful tool for any class.
- Yes, but it would be more helpful if grades were kept up to date and accurate.
- It is often slow, and it not kept particularly organized.
- Always knew all the information I needed could be found somewhere in ELMS.
- When used well by the teachers (all useful documents posted, grades posted, etc.), it is a very valuable resource for information. Using the documents posted and the lecture notes are crucial to studying for exams in the course, I've found, as everything is there to be utilized, even if you don't happen to have your book with you at all times (especially then!). If you missed something important in class, ELMS lets teachers post the information in a place accessible at any time.
- The course documents are useful references, and I like being able to see my grades. I do not use the discussion board much but I probably would if I did not know other students in the class.
- Its information resources are much more useful than both the lectures and the textbook.
- Some classes do not use ELMS very well and others are very organized and utilize it. Also, you cannot save your password to log into ELMS through Google chrome, which is painful.
- Yes, there are lots of helpful links and material all compiled together in one site to make finding everything easier.
- I like the departmental grades server and forums better.
- A great idea, but the interface is terrible. We may as well have a website with frames and an associated forum site! Computer Science is supposed to do this sort of thing better than the humanities.
- ELMS, in general, tends to be slow and difficult to navigate. Combined with mediocre organization because of old things from different semesters it is very annoying.
- I believe that for this tool to be extremely effective, we need updates for the current semester to be present. Mainly, this pertains to having the projects sent to ELMS as opposed to our emails.
- It is just as effective as a class website with lectures and other material like that. There are plenty of tools that help the class, like the discussion board.
- Having everything in one-place saves a lot of time.
- I find courses that use ELMS pretty helpful since there is a place with all of the course information, unlike other courses where I simply have a syllabus. The only reason I did not say I love it, is because sometimes the site can be rather disorganized. It can be difficult to find what you are looking for sometimes without looking through every module.
- I like how many of my courses can be integrated into one website
- I like it because of the discussion boards.
- It's very helpful for discussion purposes
- Like the forum section and having, all class assignments and materials in one place.
- For some classes it is important, for others it is not. It is only helpful when you are updating the site often, have a live discussion board, and keep important documents and grades online.
- I think ELMS is great for posting lectures slides and practice material, but I am not a big fan of the discussion board. I think some sort of separate forum would be much more useful.
- For this course, I would rather have everything be on the grades server and through email.
- It is not life changing or anything but, it is a good tool to use. I would give it an eight out of 10.
- The ability to have course information organized in one central location is extremely helpful. Also having the grade book updated makes it easy to keep track of how you are doing in the class.
- The professor actually utilizes ELMS and most of its features to make it worthwhile.
- I think ELMS course website is a helpful learning tool. However, it is great if the user name and password can be saved automatically, so I do not need to enter username and password every time I visit my courses in ELMS.
- Good to have stuff be posted and in a structured way.
- Everything is organized in a readable fashion.
- I know the instructor of this course thus, I know that if the information she had on ELMS was not here it would be on her webpage. Other teachers however if they did not have ELMS would probably email you everything and if you lost the email then you would be out of luck. So in those cases it can be helpful
- The tools for organization are not always clear, but having all (or the vast majority) of necessary online materials clumped in one place is very useful.
- ELMS has great resources, easy to check grades and online tests are awesome
- It's a useful learning tool but having information on both the instructor website and on ELMS makes it difficult to know where to go to get things
- ELMS is all right, but if all the features were used, it could be better.
- Course notes and the discussion board are really useful, as well as sample exams.

*2) Student Comments for Question 2*

The question was:
Do you prefer courses that use ELMS over other courses with a regular website?

- ELMS is annoying.
- Having a webpage is essential, though whether it is through ELMS or through a class-specific webpage does not really matter.
- It makes classes without discussion sections easier to follow.
- Especially for Computer Science classes, it does not matter because most CS teachers have their own class website and we have our own grades server so essentially all the functions of ELMS are available through the CS department specific tools.
- Only slightly better. A good website can be very good as well.
- Most other professors fail at using ELMS properly.
- ELMS lets you have a discussion board.

- I get distracted when I have to open up multiple sites for a course. It is nice to have it all in one frame where I can also access my other courses.
- Helps to have a central location for information.
- All depends on how much work the teachers put into the page. If a teacher puts lecture notes and other materials on the page then it is helpful. Other teachers of mine have said they would use ELMS and then after 2 weeks just stopped using it and resort to emails.
- It is nice to have all class websites on one page versus about 3 or 4 pages.
- Some courses benefit from ELMS more than others do. This course benefits because there are general questions about projects. Questions for this course cover more than just course material. For example, programming questions that do not directly apply to course material but are still required for projects can be asked and answered freely. Courses where questions are more material specific would be harder to pose questions to the class.
- If instructors do not use ELMS, but create their own websites with enough information it is usually okay, but often not using ELMS means not having any online source for the class.
- It is easy to bookmark classes that have individual web pages while with ELMS you always have to log in each time you access something.
- I think ELMS is a straightforward, easy to use tool that helps me access material quickly.
- In particular, I think courses with active discussion boards are the best.
- They are better, because you can get all the information you need, view your grades, etc. all in one place.
- The features I use in ELMS are also available in classes with regular web sites.
- I prefer courses that use ELMS since it has many tools such as assignments, related coursework, discussion board, and other tools and so on.
- I prefer courses that have ELMS.
- If teachers are able to maintain an updated website on their own, that is great. If not, then ELMS provides an okay way to do it.
- A website specifically designed for the course will always be better than something that tries to be generic and work for every class.
- I always know where to find information as long as the teacher updates it.
- As long as everything is in one place, it is better than not having it.
- I like being able to go to one single website and look up information for all my classes instead of having to go to several different websites.
- Some classes use ELMS well, while other barely put any materials on ELMS or use it only to distribute email material.
- Online materials are always a plus.
- My course choices are sharply limited, and it depends strongly on the professor's proactive website upkeep.
- ELMS would be even better if teachers updated the grades section so people can know how they stand in the class.
- Of course, a website is good, but it does not need to be ELMS.
- As long as, I can access my grades, but not on ELMS, I am happy.
- Courses without ELMS pages - or courses that don't utilize ELMS well - definitely lack the benefit of being able to have all the useful, relevant and important information and material easily accessible by students at any time. Being able to access information I may have missed, or misheard, or did not understand, is VERY helpful. I very much prefer classes that have this benefit.

   *3) Student Comments for Question 3*
The question was: How often do you post on the discussion boards?
- I look for questions and answers that relate to my tasks. However, I will post if I have questions or I know the answer, so it does not depend on the timeframe.
- I never post question because I prefer to ask a TA or professor through e-mail or in person if I got a question.
- I do not like to post because it shows my name!
- If I have the answer, I answer quickly, and if I have found answers to questions, I previously asked on the discussion boards.
- I did not post many posts because for programming project one and two, I did them relatively early compared to other students, and I did not look at the discussion posts afterwards.
- The discussion boards' questions are answered too slow and may not be accurate. The teacher almost never comes on it.
- The kinds of questions I have are usually ones that are not allowed to be answered...so.
- I think the discussion boards could be vastly improved. It just feels clunky and outdated. A forum layout is much better to work with. Especially with programming discussions.
- People are rude on the discussion boards at times.
- Mostly when required for classes.
- I am sort of a shy person. I answered someone's question once, but I feel as a retard for some of the questions I want to ask. It is a great idea, but a lot of times, having an attached drawing or something for an explanation would be a good idea.
- Most of time, I rely on others asking good questions and learning from their questions instead of asking my own.
- Simple technical questions may be posted. Other than that, I prefer face to face.
- I will post if I am truly confused about something.
- In general, this is a lot easier than being spammed with fifty e-mails about student questions that apply to more people.
- Definitely good when there are no office hours available.
- Would rather attend office hours.
- I only post occasionally when I have questions but not too often. Usually I just look at other people's questions.
- Semi-daily when nearing the end of projects, otherwise almost never.
- I check the message boards a lot, but most of the questions I have are already answered there and I therefore do not need to post messages.
- I rarely post my own questions because one such question usually already exists.
- Mostly look for answers, rarely post
- The discussion board is a great resource.
- Most of my classes that I have taken that have used ELMS; do not extensively use the discussion boards.
- I would post anonymously.
- I have posted before but I do not post often in THIS class. For ELMS in general, I took a class entirely based on using ELMS, and I posted every day in those discussion boards. I do not know if that is relevant or not.

   *4) Student Comments for Question 4*
The question was: If you have questions, do you prefer to go to office hours or you try the boards first?
- It is clearer, if I can talk with someone face to face, but sometimes, information in the discussion board really help me, and I can see others have the same troubles.
- Almost always, the face-to-face benefits are better than ELMS. For example, you get immediate, certain, and complete information when you sit down with a teacher and hash out a specific topic that is difficult.
- It is usually easier and more convenient to post a message to the discussion board.
- If I am nearby the TA office hours already, I will check in with them, else the discussion board is much more convenient.
- It depends on the question.
- Usually, I will ask a classmate first. If I still cannot figure it out, I might try office hours or the boards.
- It is complicated sometimes to get your point across through a message. I also like talking face to face because I never know what I can and cannot put online in terms of code.
- Better to talk in person on campus.
- I don't think a discussion board can ever be as good as talking to a TA
- ELMS is more convenient than going to office hours because it is all electronic, and some office hours are during student class time.
- Usually the answer is either not allowed to be explained, or it is too complicated to explain over discussion boards. Asking in person is usually better for me.
- I cannot really attend the office hours; I have classes at those times.
- I look for questions all over the place, and ELMS having a discussion board for that purpose definitely saves me a lot of time and effort compared to if it were not there.
- I look on all fronts.
- When I have a question that is not answered by the discussion boards, I assume it is a question only if I have and I will then go to a teacher to ask it.
- It depends on the question.
- Usually, the TAs help better than just a post on ELMS.
- The discussion board posts are the fastest method of getting the information but office hours goes more in-depth.
- If it is a simple question, I like ELMS. If it is more of a difficult question that can only be answered face to face then I will go that option.
- I prefer to go to office hours, but it is a long drive for me, so I check ELMS first.
- I prefer talking to someone if I actually have a question.
- If I have question I usually try to contact a TA first.

- I feel the forum questions are generally, "what is this" rather than the more detailed/intricate "how do I do this", which requires a face-to-face discussion.

   *5) Student Comments for Question 5*

The question was: If you posted to the board and a fellow student answered your question, do you trust his answer?

- I believe that the students who actually take the time to answer questions will know the answer correctly.
- I trust everyone!
- Unless they say something along the lines of "I'm not sure", I figure if they bother to post the answer they got it to work for themselves.
- I trust the answer of my classmates. Although there would be some wrong answers, other classmates suggest correct answer.
- That is true. My classmate might give the wrong answer since his/her posts based on his/her knowledge that is not 100% correct; however, sometimes he/she posts other sources or his/her reviews for the previous tasks is a great information.
- Since there is no anonymity, I am fairly certain none would knowingly pass on false information.
- I trust the answer if it makes sense.
- I cannot take everything for granted.
- An incorrect answer would be corrected by another student or by the instructor.
- If I can see their reasoning and it makes sense, I can trust their answer well enough. If it sounds completely farfetched or I am not sure of its accuracy, I will probably ask someone else what they think of the answer before trusting it.
- I usually just try it and see if it works, and then I will know whether I can trust the person.
- The good thing is, with a public board even if someone does make a mistake another student or the instructor can always come along to clarify.
- Also, it depends on which classmate it is.
- Sometimes I trust an answer. If I know the classmate and how smart they are or if I know the classmate has been wrong before and is not the best student then probably not.
- Usually I do unless it sounds too outrageous to be true, which has not been the case so far.
- I do trust my fellow students, but I would feel much more confident about the answer if someone from the instructional team approves.
- I trust if none has corrected the individual or has confirmed its correctness even more.
- It depends on the way in which the answer is presented and how well the answer is supported, in short, on the presentation of the answer.
- I trusted the TAs and the professor.
- I would prefer to verify a classmate's answer before putting it in practice, time permitting.
- Most of the time, I would assume they are correct unless something seems off to me.
- They are usually right and it is not hard to double check.
- Someone who can post answers that are not perfect but still they know better than I do.
- Most of the time students only post when they know the answer, but it is always assuring to know the instructional team says it is the right answer.
- Answers are tested by trying them to see if they are right. Usually they help to see another perspective.
- I would compare to my own answer.

   *6) Student Comments for Question 6*

The question was: How many posts do you read?

- I read everything, but everything I read does not actually help me at all.
- Similar questions or interesting posts.
- I typically look for things that I am having trouble with and look for similar title names.
- I read most things; in case there is, something I might have misunderstood that is clarified.
- I skipped the post saying "Please, give us an extension." :)
- I read everything on the topics I am confused about.
- A number of topics I expected to be inane have proved (marginally) useful, and I'd rather not miss information I didn't know because I didn't expect to find it there.
- I try to at least skim every post in the discussion boards - any little bit of information may be useful, and even if I do not delve into the question at the time, I may remember it later and find it useful then.
- For example, I will not look back to posts of a project after I finished it.
- point epends on how much free time I have.
- It always help to read everything
- I will read most of the posts, unless it is about specific questions that I already know the answers to them.
- In my opinion, the majority of the posts I have read in the past are made up of either not so intelligent questions or non-helpful answers, so they are not really of much value.
- I try to read all the posts everyday!
- I have them emailed to me automatically.
- Discussion board is always fun to read.
- If the post is dealing, with nothing I need then I do not read it.
- I only read posts that are relevant to the information I am looking for.
- If it is something I know will not matter, I do not read it. But in general, I like to read almost everything.
- That question was pointless.
- I usually take a peek at every post.
- I read posts posted by either the instructor or the TAs.
- Not exactly, I read all the posts that relate to my concerns, and other questions that I find helpful or can answer them.

   *7) Student Comments for Question 9*

The question was: Did you participate in this survey because of the freebie point?

- I am not going to say "false" to the previous question - who does not want an extra point? – I most likely would have taken the survey with or without being given an extra point - It is not like it is all that hard to do.
- I think ELMs is pretty good. I do wish they would upgrade the GUI, it feels like a 90's web application.
- HELL YEA!
- I participated because of the point, but I would have given feedback either way.
- Thanks for the extra point but I do feel like it was valuable to give the students input on ELMS.
- If points were not an incentive, I'm not sure I'd get this done in time because of what else is going on this week, but I hope to broaden ELMS (or an improved version of ELMS) usage in order to promote smarter learning.
- Extra points are attractive.
- It did not hurt, but I do like giving feedback on pedagogy. Student feedback is hard to get/give in the big lecture hall.
- Nevertheless, I DID answer them honestly!
- Cool survey.
- Probably would have done it eventually anyways
- I think it would be helpful if all courses have ELMS support. It's also nice for viewing grades and the syllabus and assignments.
- I do hope my feedback helps.
- Mostly, but I probably would have done so anyway if I thought it would help someone out.
- POINT!!!
- If the instructor had said that I should do it but had not offered the point, I still would have done it. If it were something that was not recommended by the instructor then I might not have done it.
- Point!
- A simple cost-benefit analysis suggests that an extra point is easily worth answering a few questions.
- While I did participated for the extra credit, I answered all questions truthfully and to the best of my ability.
- It is an incentive; incentives such as this always get more people to do surveys. It is a matter of deciding what my time is worth. Is 5 minutes of my time worth nothing to me or is it worth a few points. I like to think my time is worth something.
- However, I also do feel that this survey is important, when I started at Maryland, only one or two of my classes used ELMS as opposed to now, when I never have a class that does not use it.
- True, however I answered all questions according to how I feel and did not rush through it. : )
- I need the extra points as the instructor said.
- I need every point I can get. Don't hesitate to offer another one of these, because, I am sure my classmates would welcome it as would I.
- Nevertheless, it is also my benefit and improved the quality of the education environment.
- Hey, at least I am honest.
- It is nice to get the point, although I would have taken the survey without the

- point.
- Extra points, yay!

AUTHORS PROFILE

Abdel-Hameed A. Badawy, an ABD PhD Candidate at the Electrical and Computer Engineering Department, University of Maryland, College Park, MD. Mr. Badawy obtained a B.Sc. in Electronics Engineering with a specialization in Computing Systems and Automatic Control systems from Mansoura University, Egypt in 1996 with Distinction and Honors. He was ranked the first on his graduating class in his specialization. He obtained a graduate certificate in software development from the Egyptian Cabinet sponsored, Information Technology Institute, Giza, Egypt in 1997. He was a Teaching Assistant at the Systems and Controls department at Mansoura University from September 1996 till August 1999. He obtained his M.Sc. in Computer Engineering at the University of Maryland in August 2002 and defended his PhD proposal and advanced to candidacy in August 2006. Mr. Badawy is a Teaching Assistant Training and Development (TATD) Fellow at the Electrical and Computer Engineering Department for the academic years 2010-2011 and 2011-2012. Mr. Badawy is currently a Center for Teaching Excellence (CTE) Graduate Lilly Fellow. Mr. Badawy is a student member of IEEE, ACM. HE is also a member of the Science and Engineering Institute (SCIEI), a member of the International Association of Computer Science and Information Technology (IACSIT), an associate editor for the August 2010 issue of the Journal of Learning, reviewer for the International Journal of Advanced Computer Science and Applications (IJACSA), reviewer for the Society of Imaging Informatics in Medicine (SIIM) Journal of Digital Imaging (JDI), reviewer for the International Journal of Computer and Electrical Engineering (IJCEE), and technical program committee member for several conferences. Mr. Badawy's published research has won best student paper awards at AIPR 2010. He won the prize of excellence at the University of Maryland, Graduate Research Interaction Day 2004 (GRID). He also won a best poster award at GRID 2012. Mr. Badawy's research interests span computer architecture, compiler optimizations, machine intelligence applications in medicine including but not limited to medical imaging, and engineering and computer science Education.